# On calculation of elementary particles' masses

**Alexander G. Kyriakos**

*Saint-Petersburg State Institute of Technology,
St. Petersburg, Russia.*

*Present address: Athens, Greece, e-mail:* agkyriak@yahoo.com

Abstract. The purpose of present paper is to develop the approach to calculation of the mass spectra of elementary particles within the framework of the resonance theory of elementary particles as de Broglie waves

## 1.0. Introduction.

**1.1. Equivalence of energy and mass spectra**

As is known neither classical, nor quantum theories could explain the nature of masses of elementary particles and could deduce the numerical values of masses till now.

The basic experimental facts here are the following: 1) masses of elementary particles make the discrete spectra; 2) according to modern representations, all elementary particles are the exited states of a small set of some particles, which represent the lowest level of a spectrum of masses.

It is supposed that discreteness of spectrum of masses of elementary particles is similar to a discrete spectrum of excitation energies of atom. According to the formula $e = mc^2$, where $e$ is the particle energy, $m$ is the rest mass of particle and $c$ is the light velocity, to any rest mass corresponds the stationary level of energy.

**1.2. Energy spectra of electron in hydrogen atom as an example of a spectrum of masses**

The first calculation of energy-mass spectrum of electron in hydrogen atom has been based on the known Bohr atom theory, in which the quantization was entered by a separate postulate. This approach allows the calculation of energy spectrum of electron, but it does not reveal the reasons of quantization.

The reasons of quantization have been specified by de Broglie, who showed that elementary particles in a stationary state can be considered as standing waves, which formation conditions are the conditions of the length waves' integrality.

Below we will consider in more detail the analysis of electron motion in hydrogen atom from the point of view of the de Broglie theory.

*1.2.2. Quantization of the electron energy in hydrogen atom according to de Broglie*

In his dissertation de Broglie has shown (Broglie, de, 1924; 1925;. Andrade e Silva and Loshak, 1972), that the orbits postulated by Bohr for electron motion around a hydrogen atom nucleus can be received from the condition that the length of an orbit $L$ should contain an integer number of electron wavelengths $L = n\mathbf{l}$ (see fig. 1):

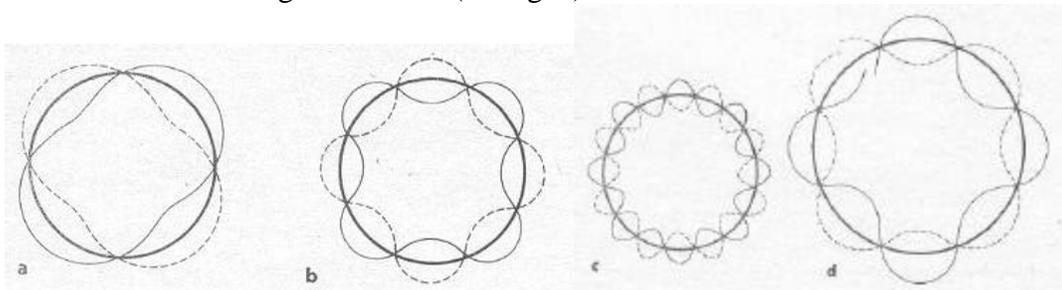

Fig. 1

where $\mathbf{\lambda} = \dfrac{h}{m\mathbf{u}} = \dfrac{2\pi\hbar}{p}$ is the particle wavelength according to de Broglie, $h, \hbar$ is the Planck constant (usual and bar), $\mathbf{u}$ is the particle velocity, $p = m\mathbf{u}$ is particle momentum. For **a**, **b** and **c** of fig.1 this condition is carried out, when $n = 2, 4$ and $8$, accordingly. In case of **d** this condition is not carried out and electron motion is unstable, that leads to self-destruction of a wave as a result of an wave interference. Mathematically the integrality condition corresponds to the requirement of unambiguity of wave function.

A similar condition also takes place for elliptic orbits (see also (Shpolski, 1951)), but this case is more complex, since the length of de Broglie wave in different points of an elliptic orbit varies because the electron speed is not constant. In this case it is necessary to use the general condition of quantization:

$$\int \dfrac{ds}{\mathbf{\lambda}} = \int_0^T \dfrac{m\mathbf{\beta}^2 c^2}{h\sqrt{1-\mathbf{\beta}^2}} dt = n, \qquad (1.1)$$

where $ds$ is the orbit length element, $T$ is the period of motion, $dt$ - time element, $\mathbf{\beta} = c/\mathbf{u}$.

Thus, calculation of electron energy levels in hydrogen atom from the wave theory point of view needs to be considered as calculation of resonance conditions of de Broglie electronic wave in the potential well (resonator) formed by an electromagnetic field of a nucleus.

From the above-stated follows that the stationarity conditions correspond to resonance conditions, which are adequate to conditions integrality of the standing waves.

Such sight at the reason of occurrence of quantum levels of electron energy also allows to calculate the last in other similar cases. For example, as an approximate model of 3-dimensional short-range potential, can be the spherical potential well of some radius $R$ (Naumov, 1984). According to de Broglie for the big circle of sphere of radius $R$, we will have:

$$2\pi R = n\mathbf{\lambda} = n\dfrac{2\pi\hbar}{p} = n\dfrac{2\pi\hbar}{\sqrt{2m\mathbf{\varepsilon}}}, \qquad (1.2)$$

From here, we receive for energy levels: $\mathbf{\varepsilon}_n = \dfrac{\hbar^2 n^2}{2mR^2}$. As we marked above (see chapter 9), the exact solution of this and other problems the Helmholtz equation for de Broglie waves gives, (on the other hand, the Schreudinger wave equation); this result differs from the approximate result only by the constant $\mathbf{\pi}^2$.

We will note one remarkable feature, which has the solution of Schreudinger equation for electron in a potential well of final depth. The solution shows (Matveev, 1989; Shiff, 1959), that in this case there is a final number of own levels of energy. Whether it is possible to distribute this conclusion to elementary particles (specifically to the leptons family), remains in doubt.

Attempts of calculation of mass-energy spectra on the basis of resonance behaviour of particles exist for a long time. We will briefly mention the most consecutive of them.

## 2.0. Present calculations of elementary particle masses

The existing calculations are based on assumptions and guesses, which cannot be proved enough within the framework of the quantum field theory.

### 2.1. Quasi-classical approaches to mass calculation

According to them the basic particle assimilates to a potential well (or, that is the same, to the resonator of the certain configuration). The spectrum of masses of particles arises, when some additional resonance particle (e.g. photon) is placed in this potential well. Characteristics of addition particle change the characteristics (mass, spin, charge, etc) of the basic particle and we can consider the last one as new particle.

One of the first attempts of quasi-classical calculation (for masses of muon and pion) belongs to Putilov (Putilov, 1964). Note that this calculation does not take into account the experimental facts, which have been found out later (e.g., the existence of a tau-lepton, the law of lepton number conservation, etc.) and it should be considered only as an example of a corresponding computational procedure.

A second, much more detailed calculation (for the big number of particles, known at that time) is stated in paper (Kenny, 1974). Here is already the theoretical substantiation of a method of calculation and are received impressing results. But calculation is made by analogy to the theory of Bohr; as a result here was used Coulomb potential well. The obtained numerical values of masses, without serious substantiation, are corresponded with masses of known particles that conduct to infringement of laws of conservation of quantum numbers, etc.

Another approach is based on the quantization rules of Bohr-Wilson-Sommerfeld. The group of scientists J.L.Ratis, F.A.Garejev and others (Ratis and Garejev, 1992; Garejev, Kazacha, Ratis, 1996; Garejev, Kazacha, Barabanov, 1998; etc) has achieved especially impressive results, using the new quantization condition for asymptotic momenta of decay products of the hadronic resonances.

**2.2. Quantum approaches to mass calculation**

The calculations, based on idea of composite particles, take place here. We will show below that this approach has near connection to the resonance theory.

As it is marked in the A. Rivero and A. Gsponer's review (Rivero and Gsponer, 2005) one of the first possible approaches to an estimation of masses of elementary particles was based on the known composite model of Nambu-Barut (Nambu, 1952; Barut, 1979). In particular, for leptons Barut has received the following formula: $m(N) = m_e \left(1 + \frac{3}{2\boldsymbol{a}} \sum_{n=0}^{n=N} n^4 \right)$, which gives satisfactory values for both heavy leptons (here $m_e$ is the electron mass, $\boldsymbol{a}$ is electromagnetic constant).

In this approach it is postulated that for calculation of masses of heavy leptons to the rest mass of electron the quantized magnetic energy $(3/2)\boldsymbol{a}^{-1} \sum_{n=0}^{n=N} n^4$ must be added, where $n$ is a new quantum number, which for for $n = 1$ gives muon mass and for $n = 2$ – taon mass $m_t = 1786{,}08$ MeV.

Recently, a similar expression had been received from other reasons. In paper (Rodriguez and Vases, 1998) for muon mass as exited state of electron (which is allocated with properties of a quark) the formula is received: $m_l = \left(1 + \frac{q_m^n}{e}\right) m_e$, where $q_m^n = \frac{3e}{2\boldsymbol{a}} n$. E.g., for muon at n = 1 turns out: $m_m = \left(1 + \frac{3}{2\boldsymbol{a}}\right) m_e = 206{,}55 m_e$. Assuming that taon is the exited state of muon, authors receive also: $m_l = \left(1 + \frac{3}{2\boldsymbol{a}}\right) m_e + \frac{q_m^n}{e} m_e = \left(1 + \frac{3}{2\boldsymbol{a}}\right) m_e + \frac{3}{2\boldsymbol{a}} n$, that at $n = 16$ gives for taon mass the value close to experimental, namely $m_t = 3494 m_e = 1781{,}9$ MeV.

Other successful empirical formula is I. Koide's formula. As authors (Rivero and Gsponer, 2005) in the review write: "on the end of year1981 I. Koide, working above some composite model of quarks, has had fortunate or unfortunate case to run into very simple correlation among three charge leptons $(m_e + m_m + m_t) = \frac{2}{3} \left(\sqrt{m_e} + \sqrt{m_m} + \sqrt{m_t}\right)^2$, which gives for mass of a tau-lepton 1777 MeV.

Below we will show that there is a successive approach, which doesn't contradict to quantum field theory, that allows to obtain strictly enough the formulas, close to the formulas of A.O. Barut, K.A Putilov., W.A Rodrigues.- J. Vaz, and also to confirm the calculation formula of Yu. L. Ratic –F.A. Gareev-et all group. (In the framework of this analysis about I. Koide formula we cannot say anything).

## 3.0. Statement of problems of calculation of masses of particles

According to the wave theory the elementary particles must be formed as wave stable formations. Hence, their characteristics (in this case, energy-mass of particles) should be calculated, proceeding from laws of the wave theory.

We suppose (Kyriakos, 2006) that
1) all particles are divided into two groups: a) absolutely stable: electron, neutrino, proton and their antiparticles; and b) metastable: all other particles.
2) the metastable elementary particles are the composite (compound) particles, appearing as superposition of an absolutely stable particle and some additive particles.
3) the metastability of compound particles is ensured by resonance conditions and by corresponding conservation laws.

As a simplest reaction of formation of a compound particle it is possible to consider the transition of electron $e^-$ in an atom from a low level to higher level of energy:

$$e^-(\boldsymbol{e}_n) + N = \boldsymbol{g} + e^-(\boldsymbol{e}_b) + N, \qquad (3.1)$$

where $\boldsymbol{e}_n > \boldsymbol{e}_b$, $\boldsymbol{e}_b$ is the base electron energy, $\boldsymbol{e}_n$ is any electron energy level, $\boldsymbol{g}$ is the photon (gamma-quantum), and $N$ means the field of the nucleus, which in this case works as a resonator. Note that the reaction of electron-positron pair production from a photon formally in the same way is also described:

$$\boldsymbol{g} + N = e^- + e^+ + N, \qquad (3.2)$$

The record (3.1) is possible to be considered as an instruction that the electron in a mass-energy state $\boldsymbol{e}_n$ "is composed" from electron in the state $\boldsymbol{e}_b$ and a photon $\boldsymbol{g}$. To this it corresponds the statement of the problem of the description of a compound particle by the solution of the wave equation for the given potential well (we will for brevity name this problem as direct problem). In particular, for reaction (3.1) such equation is the Schroedinger equation (i.e. Helmholtz equation for de Broglie waves), and for reaction (3.2) is the Dirac equation. The solution of Schroedinger equation gives spectra, which are allowed by a potential well (resonator) of system (in this case, Coulomb field of a proton). But this solution does not take into account and does not describe the concrete reason of formation of this or that level.

Another statement of the problem (an "inverse" problem of particle mass calculation) arises in case we shall write down the reaction (3.1) in the opposite direction:

$$\boldsymbol{g} + e^-(\boldsymbol{e}_b) + N = e^-(\boldsymbol{e}_n) + N, \qquad (3.3)$$

Here initial particles: $\boldsymbol{g}$-quantum and electron $e^-(\boldsymbol{e}_b)$ can be considered as the reason of occurrence of a compound particle. In this case, obviously, for calculation of energy-mass of a compound particle $e^-(\boldsymbol{e}_n)$ it is necessary to know the experimental characteristics of waves (particles), which have actually formed this level of energy.

It is not difficult to see that these problems can be considered as necessary and sufficient conditions of formation of spectra.

It is easy to understand that the problem of the first type is reduced to the solution of the non-linear wave equations of Heisenberg non-linear equation type. Unfortunately, the solution of the last, despite of a number of achievements (in particular, the existence of spectra of masses of particles has really been shown), had difficulties, which are not overcome till now. Therefore we will try to solve this problem in linear approach, taking into account the known integrated characteristics of an initial particle (in this case, of electron).

As concrete example of compound particle formation we will consider the reactions (3.1) and (3.2). In these reactions the electron behaves simultaneously both as a particle and as a wave. The following paragraphs will be devoted to the brief analysis of kinematic and wave properties of particles.

## 4.0. Kinematic characteristics of a compound particle

The above-stated reactions in a general view can be presented as follows:
$$X_0 \Leftrightarrow X_1 + X_2, \qquad (4.1)$$

where letter $X$ stands for particles; the index $i$ stands for the compound particle, and 1 and 2 – for the initial particles. For each of particles the law of conservation of energy-momentum is fair:

$$e_0^2 = c^2 p_0^2 + m_0^2 c^4, \qquad (4.2)$$

$$e_1^2 = c^2 p_1^2 + m_1^2 c^4, \qquad (4.3)$$

$$e_2^2 = c^2 p_2^2 + m_2^2 c^4, \qquad (4.4)$$

where $c$ is the speed of light, $m$ in the given chapter means the particle rest mass; the energies and momentums are defined by relativistic expressions: $e = mc^2 d$, $\vec{p} = m\vec{u}d$, where, $d = 1/\sqrt{1-b^2}$, $b = \vec{u}/c$, $\vec{u}$ is speed of a particle. Besides in the relativistic mechanics the kinetic energy is entered by the following expression:

$$e_k = mc^2(d-1) = mc^2 d - mc^2, \qquad (4.5)$$

Since $u < c$, the expressions, containing $d$, can be expanded to Maclaurin series (taken into account only 4 terms):

$$d = 1 + \left\{\frac{1}{2}b^2 + \frac{3}{8}b^4 + \frac{5}{16}b^6 + \frac{35}{128}b^8 + ...\right\}, \qquad (4.6)$$

$$bd = 0 + b + 0 + \frac{1}{2}b^3 + ..., \qquad (4.7)$$

and we can receive for energy and momentum the following expressions:

$$e = mc^2 + mc^2\left\{\frac{1}{2}b^2 + \frac{3}{8}b^4 + \frac{5}{16}b^6 + \frac{35}{128}b^8 + ...\right\}, \qquad (4.8)$$

$$e_k = e - mc^2 = mc^2\left\{\frac{1}{2}b^2 + \frac{3}{8}b^4 + \frac{5}{16}b^6 + \frac{35}{128}b^8 + ...\right\} =$$
$$= \frac{1}{2}mu^2 + \frac{3}{8}m\left(\frac{u^4}{c^2}\right) + \frac{5}{16}m\left(\frac{u^6}{c^4}\right) + \frac{35}{128}m\left(\frac{u^8}{c^6}\right) + ... \qquad (4.9)$$

$$p = mu + \frac{1}{2}m\frac{u^3}{c^2} + ..., \qquad (4.10)$$

At $b \ll 1$ we obtain from (4.8)-(4.10) as first approximation the classical expressions:

$$e \approx mc^2 + \frac{1}{2}mu^2 = e_{cl}; \quad p \approx mu = p_{cl}; \quad e_k \approx \frac{1}{2}mu^2 = e_{kcl}; \qquad (4.11)$$

where the index "$cl$" is for "*classical*"

According to laws of conservation of energy and momentum we have:
$$e_0 = e_1 + e_2, \qquad (4.12)$$

$$\vec{p}_0 = \vec{p}_1 + \vec{p}_2, \qquad (4.13)$$

Not narrowing the framework of the problem, we can consider a case when the particle is motionless; that means $\vec{p}_0 = 0$. Then from (4.2) we receive $e_0 = m_0 c^2$, and from (4.12): $\vec{p}_1 = -\vec{p}_2$. Entering a designation $|\vec{p}_1| = |\vec{p}_2| = p_r$, from (4.14) we will receive the known kinematic expression:

$$m_i c^2 = \sqrt{m_1^2 c^4 + c^2 p_r^2} + \sqrt{m_2^2 c^4 + c^2 p_r^2}, \tag{4.14}$$

Considering the particles 1 and 2 as composite parts of particle 0, we can according to N. Bohr postulate here the quantization of momentum $p_r$ and compare the solution with experimental data. But such approach does not give the explanation and substantiation of the quantization reason. For the clarification of the last we will take the de Broglie approach.

## 5.0. The wave characteristics of elementary particles

As it is known the de Broglie wave is a clearly relativistic effect, connected with relative motion of electron in relation to other bodies (in particular, to a proton). As the de Broglie analysis shows, the wave occurrence can be connected with relativistic Doppler effect, but the deep reasons of this phenomenon remain unknown for us.

According to de Broglie the particle with energy $e$ and momentum $p$ has, taking into account (4.8)-(4.10), the following frequency and wavelength:

$$\boldsymbol{n} = \frac{\boldsymbol{e}}{h} = \frac{mc^2}{h} + \frac{\boldsymbol{e}_k}{h} = \boldsymbol{n}_0 + \boldsymbol{n}(\boldsymbol{u}), \tag{5.1}$$

$$\boldsymbol{l} = \frac{h}{p} = h \bigg/ \left( m\boldsymbol{u} + \frac{1}{2} m \frac{\boldsymbol{u}^3}{c^2} + ... \right), \tag{5.2}$$

where, $\boldsymbol{n}_0 = mc^2/h$, $\boldsymbol{n}(\boldsymbol{u}) = \dfrac{1}{h}\left\{\dfrac{1}{2} m\boldsymbol{u}^2 + \dfrac{3}{8} m\left(\dfrac{\boldsymbol{u}^4}{c^2}\right) + \dfrac{5}{16} m\left(\dfrac{\boldsymbol{u}^6}{c^4}\right) + \dfrac{35}{128} m\left(\dfrac{\boldsymbol{u}^8}{c^6}\right) + ...\right\}.$

In the case $\boldsymbol{u} \ll c$ we will obtain:

$$\boldsymbol{n} = \frac{\boldsymbol{e}}{h} \approx \frac{mc^2}{h} + \frac{1}{2} m\boldsymbol{u}^2 \bigg/ h, \tag{5.3}$$

$$\boldsymbol{l} = \frac{h}{p} \approx \frac{h}{m\boldsymbol{u}}, \tag{5.4}$$

Let's analyse these expressions.

**At first**, we will note the interesting feature of de Broglie wave: it consists of infinite series of waves, which frequencies sum up arithmetically.

**Secondly**, the "base" wave exists, which high frequency $\boldsymbol{n}_0$ does not depend on the particle motion.

**Thirdly**, a number of waves exist, the frequencies of which depend on the velocity of the particle motion and correspond to separate terms of expansion $\boldsymbol{n}(\boldsymbol{u})$. The values of the frequencies of these waves are far less than the frequency $\boldsymbol{n}_0$ of the main wave, so that these waves modulate the main wave. What role do these waves separately play in nature, we do not know, but their sum defines all wave effects of particles motion: motion of electron in the atom, diffraction of electron into slots and other.

**Fourthly**, from above very serious consequence follows that: *the frequency of the "based" wave $\boldsymbol{n}_0$ of the particle defines the Compton wavelength of rest particle and its "bare" mass:*

$$\lambda_e = \frac{\hbar}{m_e c} = \frac{1}{c\boldsymbol{n}_0}.$$

**Fifthly**, at zero speed of particle motion the de Broglie *wavelength* is equal to infinity, which for classical oscillator corresponds to zero frequency of oscillation. But for de Broglie wave *frequency* in this case we obtain a value other than zero. Actually, we obtain a rather big frequency $\boldsymbol{n}_0$ of the order of $10^{15}$ Hz, which does not depend on the speed of electron motion.

It is easy to see, that in case of hydrogen atom the formation of new levels of electron energy-mass occurs not because the electron itself absorbs a photon as a resonator, but a resonator formed by a potential well of a nucleus.

It is possible to tell that in this case, nature has thought up the smart mechanism, at which to the big and constant frequency of own wave of the rest electron, corresponds the frequency of an additional wave, which changes from zero to infinity, depending on the speed of electron motion as a particle.

We will try to show now that it is possible to explain a spectrum of masses of elementary particles similarly to the spectrum of electron masses in hydrogen atom, taking into account that in this case the particle itself plays role of the resonator for additional particles.

## 6.0. To calculation of mass spectra of elementary particles

### 6.1. A direct problem

As examples of the elementary reactions of production and disintegration of elementary particles (see (Review of Particle Properties, 1994).) are:

1) reaction of electron-positron pair production $\boldsymbol{g} + N = e^- + e^+ + N$;

2) muon decay $\boldsymbol{m}^{\pm} = e^{\pm} + \boldsymbol{n} + \bar{\boldsymbol{n}}(99\%), = e^{\pm} + \boldsymbol{n} + \bar{\boldsymbol{n}} + \boldsymbol{g}(1\%)$, and taon decay $\boldsymbol{t}^{\pm} = \boldsymbol{m}^{\pm} + \boldsymbol{n}_t + \bar{\boldsymbol{n}}_m (17,37\%), \boldsymbol{m}^{\pm} + \boldsymbol{n}_t + \bar{\boldsymbol{n}}_m + \boldsymbol{g}(3,6\%)$,

where

$$m_m = 105,6 MeV, \; m_t = 1777 MeV, \; m_e = 0,51 MeV$$
$$m_{n_e} < 3ev, \; m_{n_m} < 0,19 MeV, \; m_n \approx 1 eV \quad , \quad (6.1)$$

We can consider these reactions as superposition of the twirled photons and semi-photons. . In this case muon or taon are possibly thought of as consisting of the electronic linear polarized half-wave and two neutrino circularly polarized half-waves with the opposite direction of rotation.

It is also similarly possible to consider other reactions, without the infringement of corresponding conservation laws; for example:

3) pions decay: $\boldsymbol{p}^{\pm} = \boldsymbol{m}^{\pm} + \boldsymbol{n}_m (99,98\%)$, $\boldsymbol{p}^0 = 2\boldsymbol{g}(98,79\%; = e^+ + e^- + \boldsymbol{g}(1,19\%)$, where $m_{p^0} = 134,97 MeV; \; m_{p^{\pm}} = 139,57 MeV$

We will consider a particle $X_i$ (see (4.1)) as the given resonator, and particles $X_1$, $X_2$ as the unknown waves, which satisfy to resonance conditions of this resonator.

Since the unique particle, about the sizes of which we can speak with some share of confidence, is the electron, we will consider these three reactions, in which the electron is the lowest level of a mass spectrum. (Here instead of one photon (with spin 1) in Putilov approach, we have both neutrino and antineutrino with the half spin, moving to the opposite directions; therefore for simplification of calculation of mass we will consider two neutrino as one photon).

As we have shown (Kyriakos, 2003) the electron equation can be considered not only as (in the quantum form) the Dirac equation, but in non-linear electromagnetic form as the equation of twirled semi-photon. The free Dirac electron equation is satisfied by any mass, not only the electron mass. Reasoning from this fact, let's consider the Dirac electron equation with an external field:

$$\left[ \hat{\boldsymbol{a}}_0 \left( \hat{\boldsymbol{e}} - \boldsymbol{e}_{ph} \right) + c \hat{\vec{\boldsymbol{a}}} \left( \hat{\vec{p}} - \vec{p}_{ph} \right) + \hat{\boldsymbol{b}} mc^2 \right] \boldsymbol{y} = 0, \quad (6.2)$$

We will group here the mass-energy part:

$$\left\{ \left( \hat{\boldsymbol{a}}_o \hat{\boldsymbol{e}} + c \hat{\vec{\boldsymbol{a}}} \; \hat{\vec{p}} \right) - \left[ \left( \hat{\boldsymbol{a}}_o \boldsymbol{e}_{ph} + c \hat{\vec{\boldsymbol{a}}} \; \vec{p}_{ph} \right) + \hat{\boldsymbol{b}} \; mc^2 \right] \right\} \boldsymbol{y} = 0, \quad (6.3)$$

In this case the solution gives $\boldsymbol{y}$ function, which accepts a discrete number of values, so that to $\boldsymbol{y}^2$ there should correspond a discrete series of electron energy-masses:

$$m(n) = \left[\left(\hat{\boldsymbol{a}}_o \boldsymbol{e}_{ph} + c\hat{\vec{\boldsymbol{a}}} \vec{\boldsymbol{p}}_{ph}\right) - \hat{\boldsymbol{b}} mc^2\right] = \hat{\boldsymbol{b}}(m_e + m_{ad}) c^2, \tag{6.4}$$

where $m_{ad} = m_{ad}(n)$ is the additional mass, accepting a discrete number of values depending on $n = 1,2,3,...$. Obviously, this additional energy-mass corresponds to energy of particles, initiating the transition of electron from the level of energy $m_e c^2$ to the level of energy $(m_e + m_{ad}) c^2$.

We will try now to transform a mass term of Dirac equation so that it included the wavelengths of these particles, when they are in rest (or in other words, the sizes of particles (Kyriakos, 2006)). Substituting (6.4) in (6.2) and taking into account $\hat{\boldsymbol{e}} = i\hbar \partial/\partial t$, $\hat{\vec{p}} = -i\hbar\vec{\nabla}$, we will receive:

$$\left[\left(\hat{\boldsymbol{a}}_o \frac{\P}{\P t} - c\hat{\vec{\boldsymbol{a}}}\vec{\nabla}\right) + i\hat{\boldsymbol{b}} c \frac{(m_e + m_{ad})c}{\hbar}\right]\boldsymbol{y} = 0, \tag{6.5}$$

It is possible to present the mass term in (6.5) (without coefficient $i\hat{\boldsymbol{b}} c$) as follows:

$$\frac{(m_e + m_{ad})c}{\hbar} = \frac{m_e c}{\hbar} + \frac{m_{ad} c}{\hbar} = \frac{1}{\lambdabar_e} + \frac{1}{\lambdabar_{ad}}, \tag{6.6}$$

where $\lambdabar_e$, $\lambdabar_{ad}$ are both the Compton waves' lengths (bar) of the electron and of the additional mass, accordingly (where by definition $\lambdabar_C = \hbar/mc = \boldsymbol{l}_C/2\boldsymbol{p} mc$; note that the value $\boldsymbol{l}_C = h/mc$ also is refered to as Compton wave length). Since the basic wave contains an integer number of the additional waves (i.e. the basic wave and additional waves should be commensurable), they should satisfy the following condition of wave quantization:

$$\boldsymbol{l}_{ad} = \boldsymbol{k}\frac{\boldsymbol{l}_e}{n} \text{ or } \lambdabar_{ad} = \boldsymbol{k}\frac{\lambdabar_e}{n}, \tag{6.7}$$

where $\boldsymbol{k}$ is the number, describing a condition of occurrence of a resonance (longitudinal, cross-sectional resonance, etc.); $n = 1,2,3,...$ is an integer (quantum number). In case of propagation of a wave along the circle (as in the above problem (1.2)) we have $\boldsymbol{k} = 2\boldsymbol{p}$. In case of wave propagation along the sphere radius $\boldsymbol{k} = 4$, along the cylinder radius $\boldsymbol{k} = 2$. It is possible to assume, that generally in various configurations of particles and fields the constant can also accept other values.

Thus, for mass term in Dirac equation (i.e. for mass of a complex elementary particle) we receive:

$$\frac{c}{\hbar}m_{ep} = \frac{(m_e + m_{ad})c}{\hbar} = \frac{1}{\lambdabar_e} + \frac{1}{\lambdabar_{ad}} = \frac{1}{\lambdabar_e}\left(1 + \frac{\lambdabar_e}{\lambdabar_{ad}}\right), \tag{6.8}$$

Since the value $\boldsymbol{a} = e^2/\hbar c = r_0/\lambdabar_e \approx 1/137$ represents an electromagnetic constant, we have $\lambdabar_e = r_0/\boldsymbol{a}$ (where $r_0 = e^2/m_e c^2$ is the classical electron radius). Taking this into account, from (6.8) we will receive:

$$\frac{c}{\hbar}m_{ep} = \frac{1}{\lambdabar_e}\left(1 + \frac{\lambdabar_e}{\lambdabar_{ad}}\right) = \frac{\boldsymbol{a}}{r_0}\left(1 + \frac{r_0}{\boldsymbol{a}\lambdabar_{ad}}\right), \tag{6.9}$$

As we have shown (Kyriakos, 2004a), the "bare" size of electron corresponds to Compton wave length and at polarization in physical vacuum, decreases in $1/\boldsymbol{a} \approx 137$ times. Thus, taking into account the polarization of vacuum, instead of (6.7), we should write down:

$$\boldsymbol{l}_{ad} = \boldsymbol{k}\frac{r_0}{n} \text{ èëè or } \lambdabar_{ad} = \boldsymbol{k}\frac{r_0}{2\boldsymbol{p} n}, \tag{6.10}$$

From here $\frac{r_0}{\lambdabar_{ad}} = \frac{2\boldsymbol{p}}{\boldsymbol{k}}n$, which by substitution in the formula (6.9), gives the formula for mass of a complex particle:

$$m_{ep} = \left(1 + \frac{2\boldsymbol{p}}{\boldsymbol{k}a}n\right)m_e \approx \left(1 + \frac{2\boldsymbol{p}}{\boldsymbol{k}}137 \cdot n\right)m_e, \qquad (6.11)$$

1) for $n = 0$ we receive a trivial case of mass electronà $m_{ep} = m_e$
2) for $\boldsymbol{k} = 4$, $n = 1$ we receive $m_{ep} = 110{,}2$ MeV (that corresponds $m_m = 105{,}6$ MeV)
3) for $\boldsymbol{k} = 4$, $n = 16$ we receive $m_{ep} = 1755$ MeV (that corresponds $m_t = 1777$ MeV)
4) for $\boldsymbol{k} = \boldsymbol{p}$, $n = 1$ we receive $m_{ep} = 140{,}25$ MeV (that corresponds $m_{p^\pm} = 139{,}57$ MeV).

Results 3) and 4) are close to the results received from Putilov (Putilov, 1964)..

Despite the satisfactory concurrence of calculated and experimental values of masses of known particles, it would not be necessary to make hasty conclusions, considering an opportunity of casual concurrences. But the sequence of a theoretical conclusion of the mass formula makes the casual concurrence improbable. It is also confirmed with calculations of particle masses according to the inverse problem.

**6.2. Inverse problem.**

We will consider here a particle $X_0$ of the reaction (4.1) as the unknown resonator, and particles $X_1$, $X_2$ as the initial waves, which create this resonator.

In each resonator there are at least three sizes $L_j$ ($j = 1,2,3$), which define lengths of resonance waves. In conformity with requirements of occurrence of standing waves in the resonator, we can write down, at least, three resonance conditions:

$$L_j / \boldsymbol{l}_i = \boldsymbol{k} n_{ij}, \qquad (6.12),$$

where $i$ is number of a particle defining quantity of particles, participating in synthesis (in our example $i = 1, 2$), $n_{ij} = 1,2,3,...$ is an integer, $\boldsymbol{k}_j$ is the constant dimensionless coefficient, defining resonance conditions. Since according to de Broglie $\boldsymbol{l}_i = h/p_i$, the formula (6.12) can be rewritten in the form of:

$$p_{ij} \cdot L_j = \boldsymbol{k}_j h n_{ij}, \qquad (6.13)$$

From here at $n_{ij} = 1$ we receive the following condition for lowest states of a particle $X_0$:

$$p_{ij0} = \boldsymbol{k}_j \cdot h / L_j = const, \qquad (6.14)$$

Then for any other "exited" state of a particle $X_0$ we have:

$$p_{ij} = \boldsymbol{k}_j \frac{h}{L_j} n_{ij} = p_{ij0} \cdot n_{ij}, \qquad (6.15)$$

In case of two fussion particles ($i = 2$) we have for lowest momentums and quantum numbers values, which depend only on the index $j$, namely $p_{1j} = p_{2j} = p_{j0}$ and $n_j$. Thus, knowing only one number for the given resonance, we can calculate by (4.16) the masses for different $j$.

In works of group of Ratis, Yu.L., Garejev, F.A. et all (Ratis and Garejev, 1992; Garejev, Kazacha et al, 1998;, etc.) values $p_{j0}$ for the big group of hadron resonances, which give encouraging acknowledgement to our calculations, were selected.

In case if the quantity of particles $X_i$ is more than 2 (i.e. $i > 2$) it is necessary to have additional correlations between $p_{ij}$ for different $i$ for the calculation of spectra. Probably, the correlations, received in paragraph 4, can be useful in this case.

As sample, we present from the paper (Garejev, Kazacha, etc., 1998) the results of calculation of several resonances (see below the appendix). (Note that the paper (Garejev, Barabanov, etc., 1997) contains the analysis of some hundreds resonances, corresponding to the above conditions).

## The appendix (the data is taken from (Garejev, Kazacha, etc., 1998)):

**Table 1.** Invariant masses of the resonances which are decay along binary channels with momentums, which are multiple to 29,7918 MeV/c: Pn = n x 29,7918 MeV/c

| Resonanses | Decay chanels | $P_{exp}$ | n | $P_{exp}/n$ | $M_{exp}$ | $M_{th}$ | $\Delta M$ |
|---|---|---|---|---|---|---|---|
| $\pi^{\pm}$ | $\mu^{\pm}\nu_{\mu}$ | 29,79 | 1 | 29,79 | 139,56 | 139,56 | -- |
| $r(770)$ | $\pi^{\pm}\pi^{\mp}$ | 358 | 12 | 29,83 | 768,5 | 767,56 | 0,94 |
| $f_2(1810)$ | $\pi^{\pm}\pi^{\mp}$ | 896,70 | 30 | 29,89 | 1815 | 1809,17 | 2,17 |
| $r_5(2350)$ | $p\bar{p}$ | 714,75 | 24 | 29,87 | 2359 | 2359,31 | 0,31 |
| X(2850) | $p\bar{K}^0$ | 171,08 | 6 | 28,51 | 2850 | 2850,0 | 0,0 |

**Table 2.** Invariant masses of the resonances, which decay along binary channels with momentums, which are multiple to 26,1299 MeV/c: Pn=n x 26,1299 MeV/c

| Resonanses | Decay chanels | $P_{exp}$ | n | $P_{exp}/n$ | $M_{exp}$ | $M_{th}$ | |
|---|---|---|---|---|---|---|---|
| $\pi^0$ | $\mu^{\pm}e^{\mp}$ | 26,12 | 1 | 26,12 | 134,97 | 134,97 | -- |
| $D^0$ | $\bar{K}^0 f_0(980)$ | 549 | 21 | 26,14 | 1864,5 | 1863,97 | 0,53 |
| $D^{\pm}$ | $\bar{K}^0 \pi^{\pm}$ | 862 | 33 | 26,12 | 1869,3 | 1869,11 | 0,19 |
| $\Lambda_c^+$ | $\Xi(1530)^0 K^+$ | 471 | 18 | 26,17 | 2284,9 | 2284,25 | 0,65 |
| $B^0$ | $e^+ e^-$ | 2639 | 101 | 26,13 | 5279,2 | 5278,24 | 0,96 |